\documentclass{ws-procs9x6}

\begin{document}

\title{
Lattice QCD Study for the Interquark Force in Three-Quark
 and Multi-Quark Systems}

\author{H.~Suganuma$^{\lowercase{a}}$, T.~T.~Takahashi$^{\lowercase{b}}$, 
F.~Okiharu$^{\lowercase{c}}$ and H.~Ichie$^{\lowercase{a}}$
\vspace{0.5cm}
}

\address{a) Faculty of Science, Tokyo Institute of Technology,\\
Ohokayama 2-12-1, Meguro, Tokyo 152-8551, Japan\\
suganuma@th.phys.titech.ac.jp}

\address{b) Yukawa Institute for Theoretical Physics, Kyoto University,\\
Kitashirakawa-Oiwake, Sakyo, Kyoto 606-8502, Japan}

\address{c) Department of Physics, Nihon University,\\
Kanda-Surugadai 1-8, Chiyoda, Tokyo 101, Japan}

\maketitle

\abstracts{
We study the three-quark and multi-quark potentials in SU(3) lattice QCD. 
From the accurate calculation for 
more than 300 different patterns of 3Q systems, the static ground-state 3Q potential 
$V_{\rm 3Q}^{\rm g.s.}$ is found to be well described 
by the Coulomb plus Y-type linear potential (Y-Ansatz) within 1\%-level deviation.
As a clear evidence for Y-Ansatz, 
Y-type flux-tube formation is actually observed on the lattice in maximally-Abelian projected QCD.
For about 100 patterns of 3Q systems, 
we perform the accurate calculation for the 1st excited-state 3Q potential $V_{\rm 3Q}^{\rm e.s.}$
by diagonalizing the QCD Hamiltonian in the presence of three quarks, and find 
a large gluonic-excitation energy $\Delta E_{\rm 3Q} \equiv V_{\rm 3Q}^{\rm e.s.}-V_{\rm 3Q}^{\rm g.s.}$ 
of about 1 GeV, which gives a physical reason of  
the success of the quark model.
$\Delta E_{\rm 3Q}$ is found to be reproduced by the ``inverse Mercedes Ansatz'', 
which indicates a complicated bulk excitation for the gluonic-excitation mode. 
We study also the tetra-quark and the penta-quark potentials in lattice QCD, 
and find that they are well described by 
the OGE Coulomb plus multi-Y type linear potential, which supports the flux-tube picture even for the multi-quarks.
Finally, the narrow decay width of penta-quark baryons is discussed in terms of the QCD string theory.
}

\section{Introduction}

Quantum chromodynamics (QCD), the SU(3) gauge theory,  
was first proposed by Yoichiro Nambu\cite{N66} in 1966 as a candidate for 
the fundamental theory of the strong interaction, 
just after the introduction of the ``new" quantum number, ``color".\cite{HN65} 
In spite of its simple form, QCD creates thousands of hadrons and leads to various interesting nonperturbative phenomena 
such as color confinement\cite{N6970,N74,conf2003} and dynamical chiral-symmetry breaking.\cite{NJL61}
Even at present, it is very difficult to deal with QCD analytically due to its strong-coupling nature in the infrared region.
Instead, lattice QCD has been applied as the direct numerical analysis for nonperturbative QCD.

In 1979, the first application\cite{C7980} of lattice QCD Monte Carlo simulations was done 
for the inter-quark potential between a quark and an antiquark using the Wilson loop.
Since then, the study of the inter-quark force has been one of the important issues in lattice QCD.\cite{R97}
Actually, in hadron physics, the inter-quark force can be regarded as an elementary quantity 
to connect the ``quark world" to the ``hadron world", and plays an important role to hadron properties. 

In this paper, we perform the detailed and high-precision analyses 
for the inter-quark forces in the three-quark and the multi-quark systems with SU(3) 
lattice QCD,\cite{TMNS99,TMNS01,TSNM02,TS03,TS04,STI04,STOI04,OST04,OST04p}
and try to extract the proper picture of hadrons.

\section{The Three-Quark Potential in Lattice QCD}

In general, the three-body force is regarded as a residual interaction in most fields in physics.
In QCD, however, the three-body force among three quarks is 
a ``primary" force reflecting the SU(3) gauge symmetry.
In fact, the three-quark (3Q) potential is directly responsible 
for the structure and properties of baryons, 
similar to the relevant role of the Q$\bar{\rm Q}$ potential for meson properties. 
Furthermore, the 3Q potential is the key quantity to clarify the quark confinement in baryons.
However, in contrast to the Q$\bar{\rm Q}$ potential,\cite{R97}
there were only a few pioneering lattice studies\cite{SW8486} done in 80's  
for the 3Q potential before our study in 1999,\cite{TMNS99}
in spite of its importance in hadron physics. 

As for the functional form of the inter-quark potential, we note two theoretical arguments 
at short and long distance limits.
\begin{itemize}
\item[1.]
At the short distance, perturbative QCD is applicable,
and therefore inter-quark potential is expressed as the sum of the two-body OGE Coulomb potential. 
\item[2.]
At the long distance, the strong-coupling expansion of QCD is plausible, and it 
leads to the flux-tube picture.\cite{KS75CKP83}
\end{itemize}
Then, we theoretically conjecture the functional form of the inter-quark potential 
as the sum of OGE Coulomb potentials and the linear potential based on the flux-tube picture,
\begin{eqnarray}
V=\frac{g^2}{4\pi}\sum_{i<j}\frac{T^a_iT^a_j}{|{\bf r}_i-{\bf r}_j|}+\sigma L_{\rm min}+C,
\end{eqnarray}
where $L_{\rm min}$ is the minimal value of the total length of the flux-tube linking the static quarks.
Of course, it is highly nontrivial that these simple arguments on UV and IR limits of QCD hold for the intermediate region. 
Nevertheless, the Q$\bar {\rm Q}$ potential $V_{\rm Q\bar Q}(r)$ is well described with this form 
as\cite{R97,TMNS01,TSNM02} 
\begin{eqnarray}
V_{\rm Q \bar Q}(r)=-\frac{A_{\rm Q\bar Q}}{r}+\sigma_{\rm Q \bar Q}r+C_{\rm Q\bar Q}.
\end{eqnarray}
For the 3Q system, there appears a junction which connects the three flux-tubes from the three quarks, 
and Y-type flux-tube formation is expected.\cite{TMNS01,TSNM02,KS75CKP83,FRS91,BPV95}
Therefore, the (ground-state) 3Q potential is expected to be 
the Coulomb plus Y-type linear potential, i.e., Y-Ansatz,
\begin{eqnarray}
V_{\rm 3Q}^{\rm g.s.}=-A_{\rm 3Q}\sum_{i<j}\frac1{|{\bf r}_i-{\bf r}_j|}+
\sigma_{\rm 3Q}L_{\rm min}+C_{\rm 3Q},
\end{eqnarray}
where $L_{\rm min}$ is the length of the Y-shaped flux-tube.

For more than 300 different patterns of spatially-fixed 3Q systems, 
we calculate the ground-state 3Q potential $V_{\rm 3Q}^{\rm g.s.}$ 
from the 3Q Wilson loop $W_{\rm 3Q}$ 
using SU(3) lattice QCD\cite{TMNS01,TSNM02,TS03,TS04} 
with the standard plaquette action at the quenched level 
on various lattices, i.e.,  
($\beta$=5.7, $12^3\times 24$),
($\beta$=5.8, $16^3\times 32$), 
($\beta$=6.0, $16^3\times 32$) and 
($\beta=6.2$, $24^4$).
For the accurate measurement, we construct the ground-state-dominant  
3Q operator using the smearing method.
Note that the lattice QCD calculation is completely independent of any Ansatz for the potential form.

To conclude, we find that the static ground-state 3Q potential $V_{\rm 3Q}^{\rm g.s.}$
is well described by the Coulomb plus Y-type linear potential (Y-Ansatz)  
within 1\%-level deviation.\cite{TMNS01,TSNM02}
To demonstrate this, we show in Fig.1(a) the 3Q confinement potential $V_{\rm 3Q}^{\rm conf}$, 
i.e., the 3Q potential subtracted by the Coulomb part, 
plotted against the Y-shaped flux-tube length $L_{\rm min}$.
For each $\beta$, clear linear correspondence is found between the 3Q confinement potential 
$V_{\rm 3Q}^{\rm conf}$ and $L_{\rm min}$, 
which indicates Y-Ansatz for the 3Q potential. 

Recently, as a clear evidence for Y-Ansatz, 
Y-type flux-tube formation is actually observed 
in maximally-Abelian (MA) projected lattice QCD 
from the measurement of the action density  
in the spatially-fixed 3Q system.\cite{STI04,IBSS03} (See Figs.1 (b) and (c).) 
In this way, together with recent several analytical studies,\cite{KS03,C04}
Y-Ansatz for the static 3Q potential seems to be almost settled. 

\begin{figure}[h]
\begin{center}
\includegraphics[height=6cm]{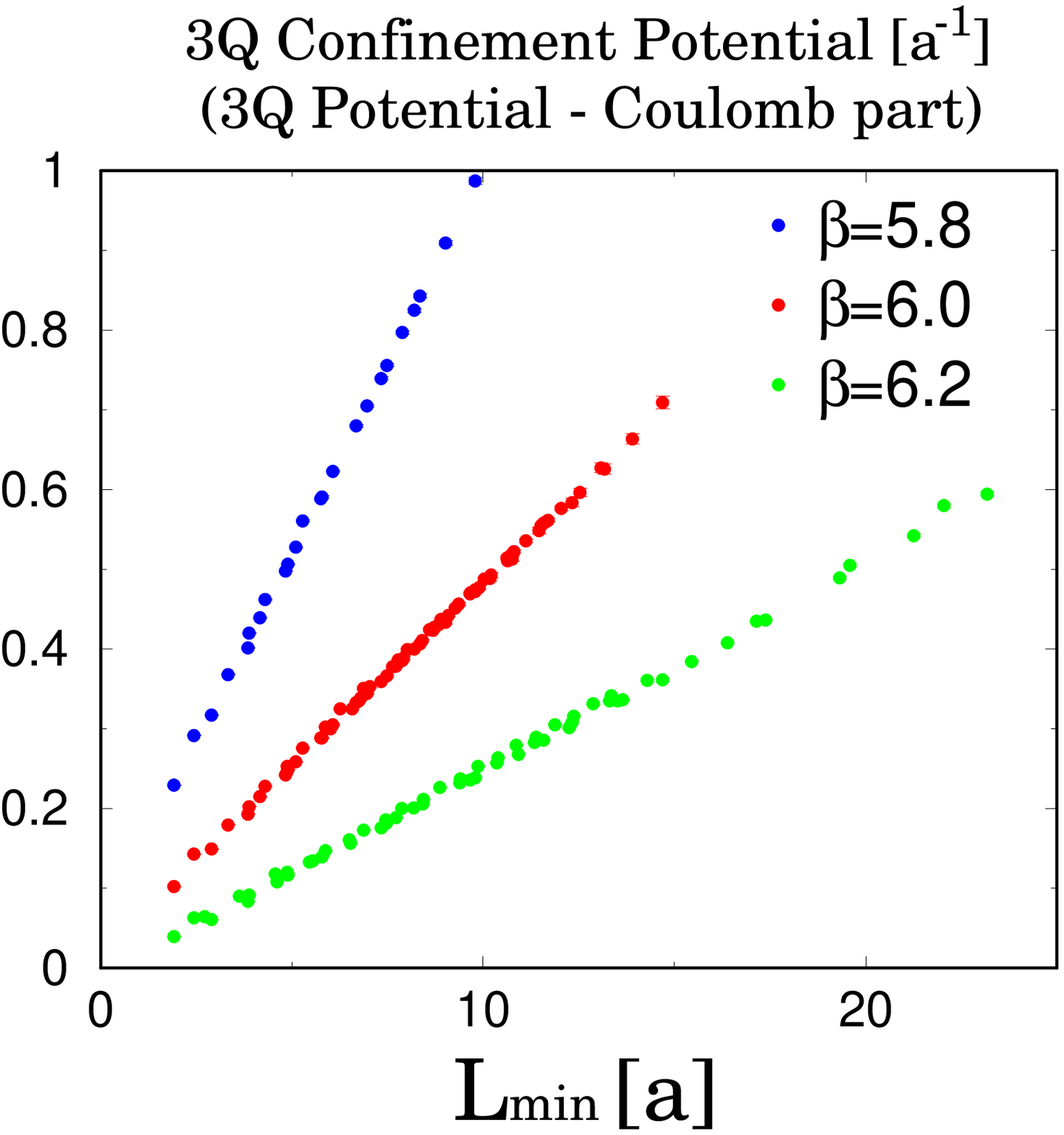}
\includegraphics[height=4.5cm]{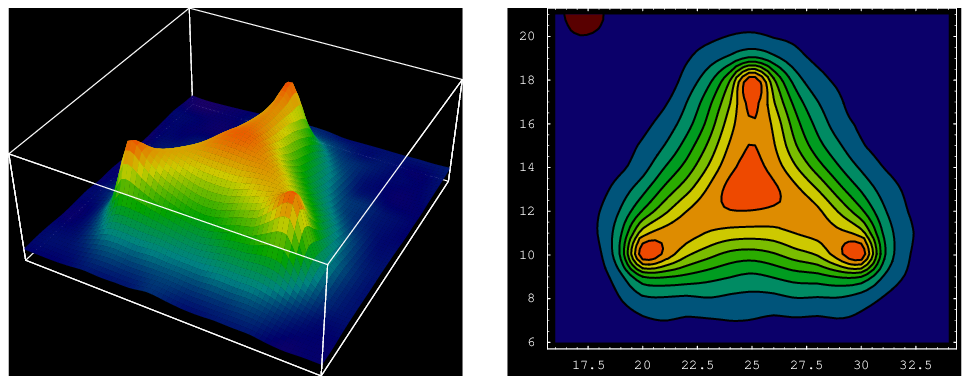}
\caption{
(a) The 3Q confinement potential $V_{\rm 3Q}^{\rm conf}$, 
i.e., the 3Q potential subtracted by the Coulomb part, 
plotted against 
the total flux-tube length $L_{\rm min}$ of Y-Ansatz
in the lattice unit.
(b) \& (c) 
The lattice QCD result for Y-type flux-tube formation 
in the spatially-fixed 3Q system 
in MA projected QCD.
The distance between the junction and each quark is about 0.5 fm.
}
\end{center}
\end{figure}

\section{The Gluonic Excitation in the 3Q System}

In this section, 
we study the excited-state 3Q potential and the gluonic excitation in the 3Q system 
using lattice QCD.\cite{TS03,TS04}
The excited-state 3Q potential $V_{\rm 3Q}^{\rm e.s.}$ is 
the energy of the excited state in the static 3Q system.
The energy difference $\Delta E_{\rm 3Q} \equiv V_{\rm 3Q}^{\rm e.s.}-V_{\rm 3Q}^{\rm g.s.}$ 
between $V_{\rm 3Q}^{\rm g.s.}$ and $V_{\rm 3Q}^{\rm e.s.}$ 
is called as the gluonic-excitation energy, and 
physically means the excitation energy of the gluon-field configuration 
in the static 3Q system.
In hadron physics, the gluonic excitation is one of the interesting phenomena 
beyond the quark model, and relates to the hybrid hadrons such as $q\bar qG$ and $qqqG$ in the valence picture. 

For about 100 different patterns of 3Q systems, 
we calculate the excited-state potential 
in SU(3) lattice QCD with $16^3\times 32$ at $\beta$=5.8 and 6.0 at the quenched level 
by diagonalizing the QCD Hamiltonian in the presence of three quarks. 
In Fig.2, we show the 1st excited-state 3Q potential $V_{\rm 3Q}^{\rm e.s.}$ and 
the ground-state potential $V_{\rm 3Q}^{\rm g.s.}$.
The gluonic excitation energy $\Delta E_{\rm 3Q} \equiv V_{\rm 3Q}^{\rm e.s.}-V_{\rm 3Q}^{\rm g.s.}$ 
in the 3Q system is found to be about 1GeV 
in the hadronic scale as $0.5{\rm fm} \le L_{\rm min} \le1.5{\rm fm}$.
Note that the gluonic excitation energy of about 1GeV is rather large compared with 
the excitation energies of the quark origin. 
This result predicts that the lowest hybrid baryon $qqqG$ has a large mass of about 2 GeV.

\begin{figure}[h]
\begin{center}
\includegraphics[height=4.5cm]{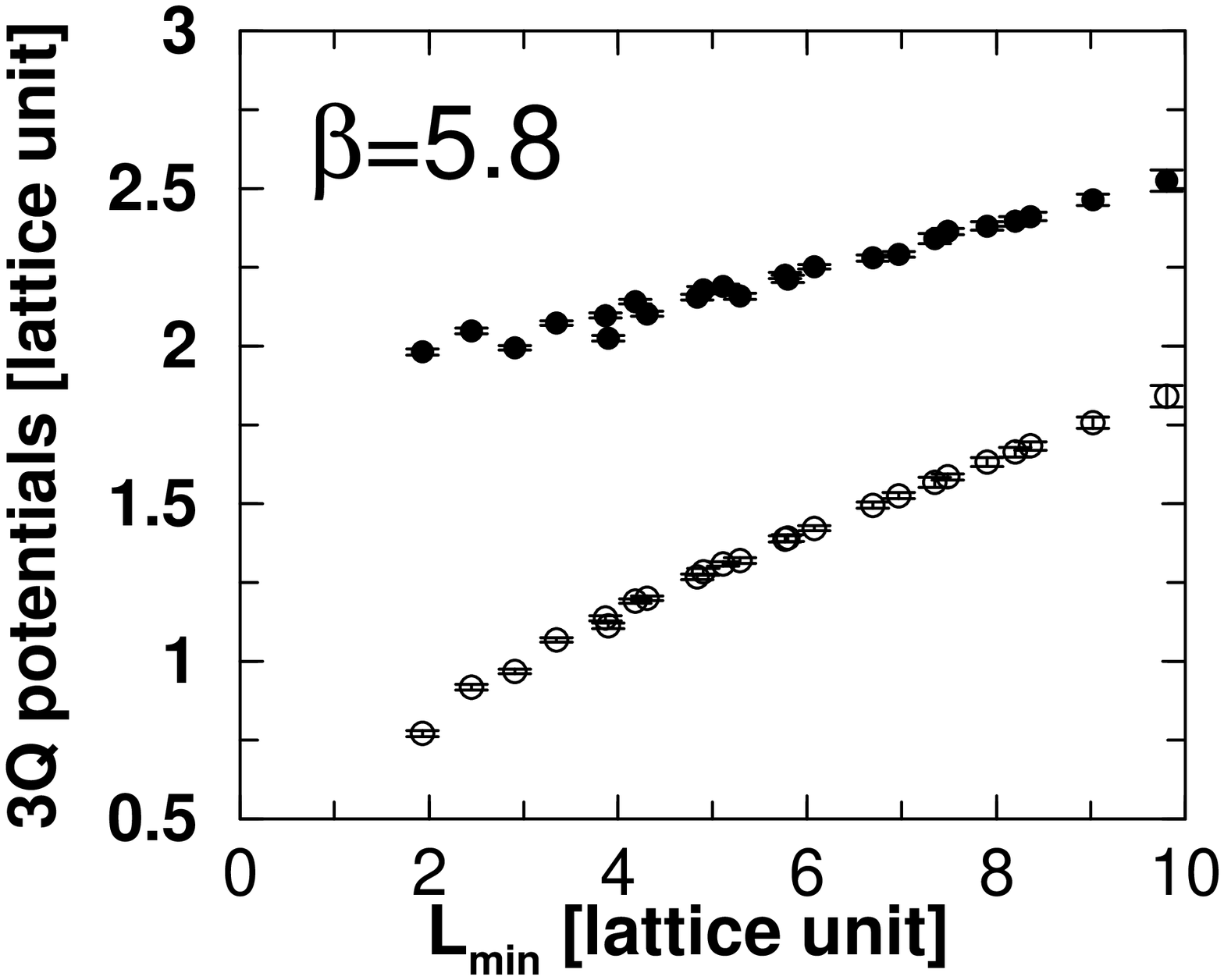}
\includegraphics[height=4.5cm]{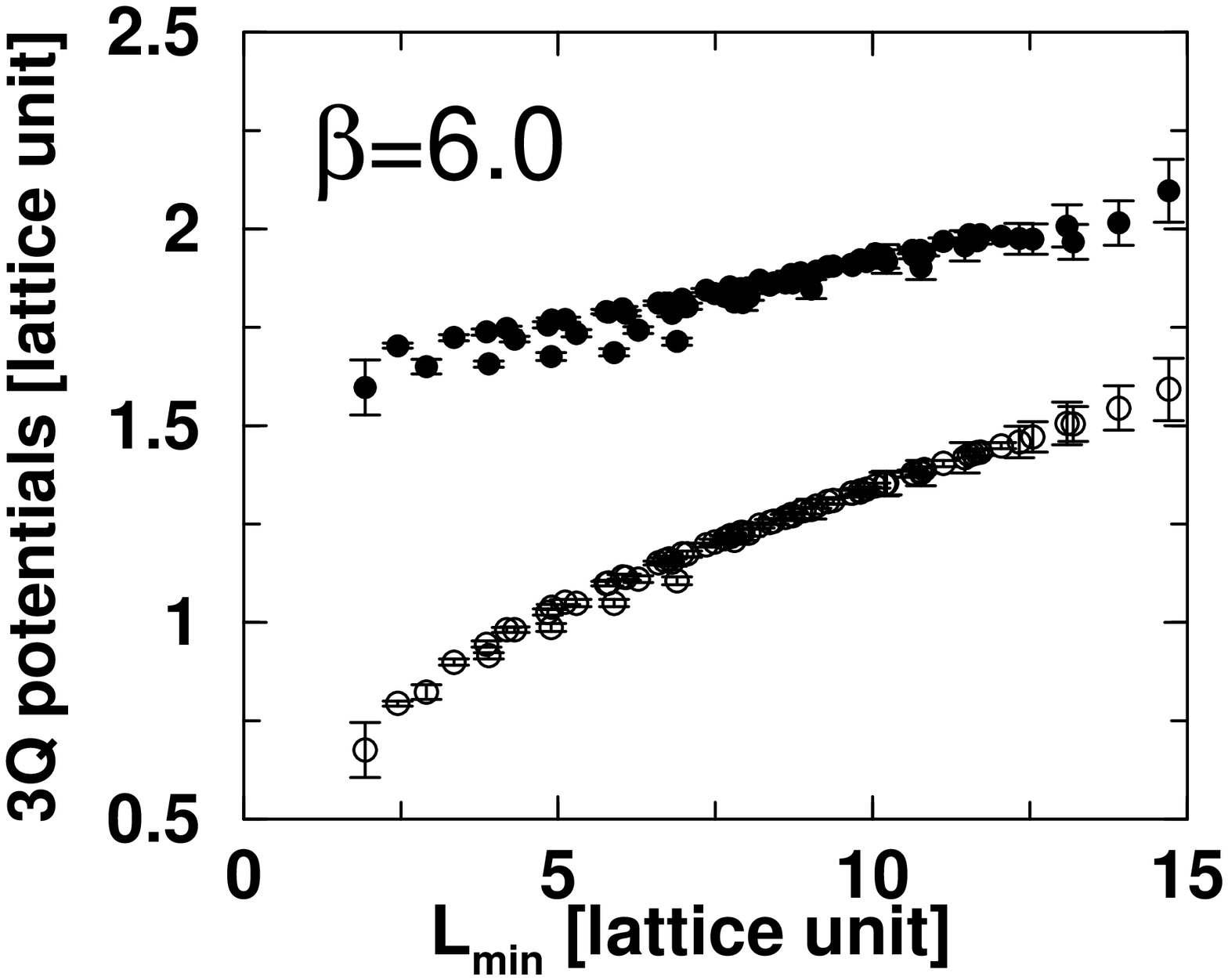}
\caption{
The 1st excited-state 3Q potential 
$V_{\rm 3Q}^{\rm e.s.}$ and 
the ground-state 3Q potential $V_{\rm 3Q}^{\rm g.s.}$.
The lattice results at $\beta=5.8$ and $\beta=6.0$ well coincide 
except for an irrelevant overall constant. 
The gluonic excitation energy 
$\Delta E_{\rm 3Q} \equiv V_{\rm 3Q}^{\rm e.s.}-V_{\rm 3Q}^{\rm g.s.}$ 
is about 1GeV in the hadronic scale 
as $0.5{\rm fm} \le L_{\rm min} \le 1.5{\rm fm}$.
}
\end{center}
\end{figure}

\vspace{-0.7cm}

\subsection{Inverse Mercedes Ansatz for the Gluonic Excitation in 3Q Systems}

Next, we investigate the functional form of $\Delta E_{\rm 3Q} \equiv V_{\rm 3Q}^{\rm e.s.}-V_{\rm 3Q}^{\rm g.s.}$, 
where the Coulomb part is expected to be canceled between $V_{\rm 3Q}^{\rm g.s.}$ and $V_{\rm 3Q}^{\rm e.s.}$. 
After some trials, as shown in Fig.3, we find that the lattice data of the gluonic excitation energy 
$\Delta E_{\rm 3Q} \equiv V_{\rm 3Q}^{\rm e.s.}-V_{\rm 3Q}^{\rm g.s.}$ 
are relatively well reproduced by the ``inverse Mercedes Ansatz'',\cite{TS04} 
\begin{eqnarray}
\Delta E_{\rm 3Q} &=&\frac{K}{L_{\rm\bar Y}}+G, \\
L_{\rm\bar Y} &\equiv& {\sum_{i=1}^{3}\sqrt{x_i^2-\xi x_i+\xi^2}}
=\frac12 \sum_{i\ne j}\overline{{\rm P}_i{\rm Q}_j}
\quad
(x_i \equiv \overline{\rm PQ}_i, \ \xi \equiv \overline{\rm PP}_i), \nonumber
\end{eqnarray}
where $L_{\rm \bar Y}$ denotes the ``modified Y-length" defined by the half perimeter of the ``Mercedes form"  
as shown in Fig.3(a).
As for ($K$, $G$, $\xi$), we find 
($K \simeq 1.43$, $G\simeq$ 0.77 GeV, $\xi \simeq$ 0.116 fm) at $\beta=5.8$, and 
($K \simeq 1.35$, $G \simeq$ 0.85 GeV, $\xi \simeq$ 0.103 fm) at $\beta=6.0$.

The inverse Mercedes Ansatz indicates that the gluonic-excitation mode is realized 
as a complicated bulk excitation of the whole 3Q system. 

\begin{figure}[h]
\begin{center}
\includegraphics[height=4.5cm]{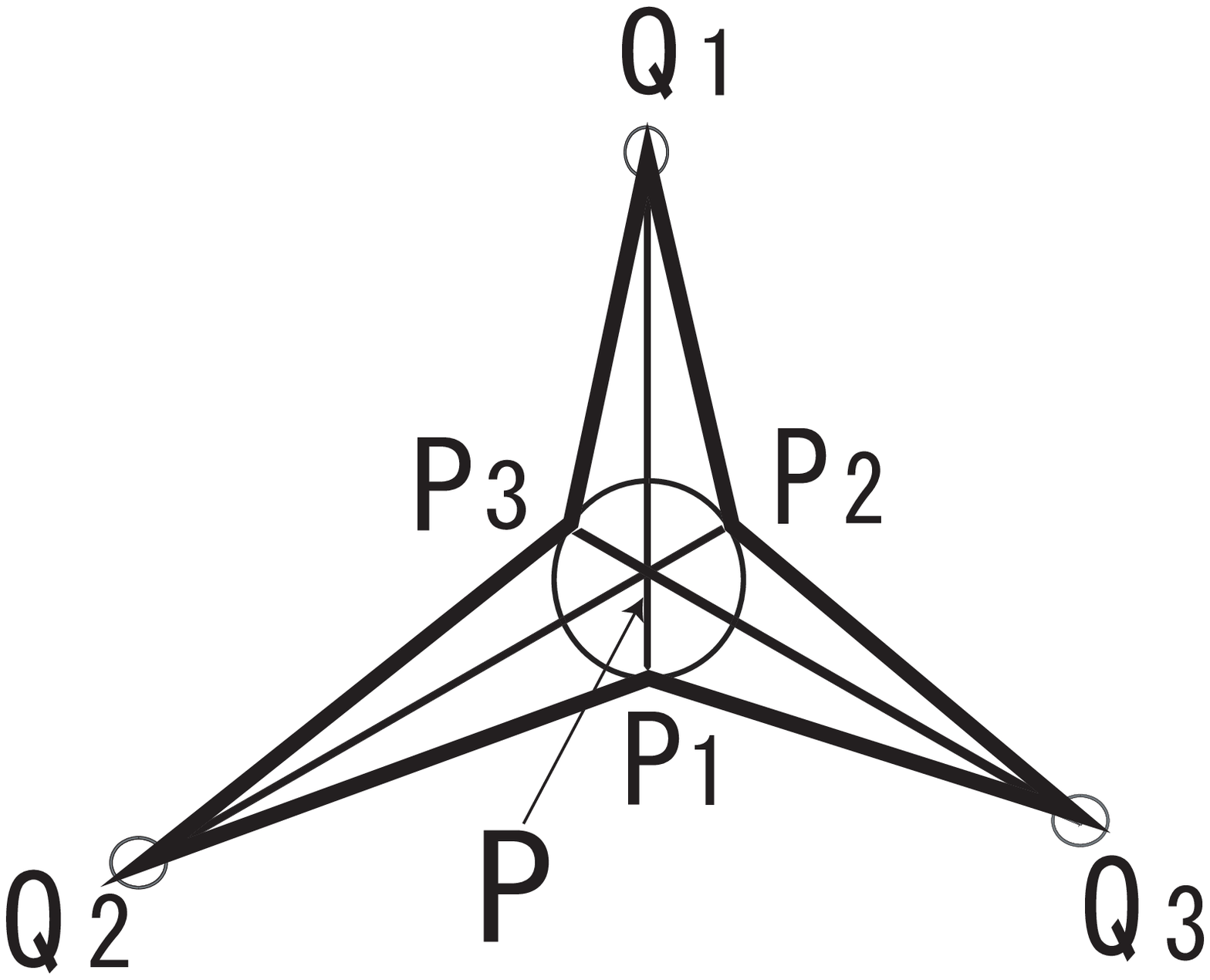}\\
\includegraphics[height=4.5cm]{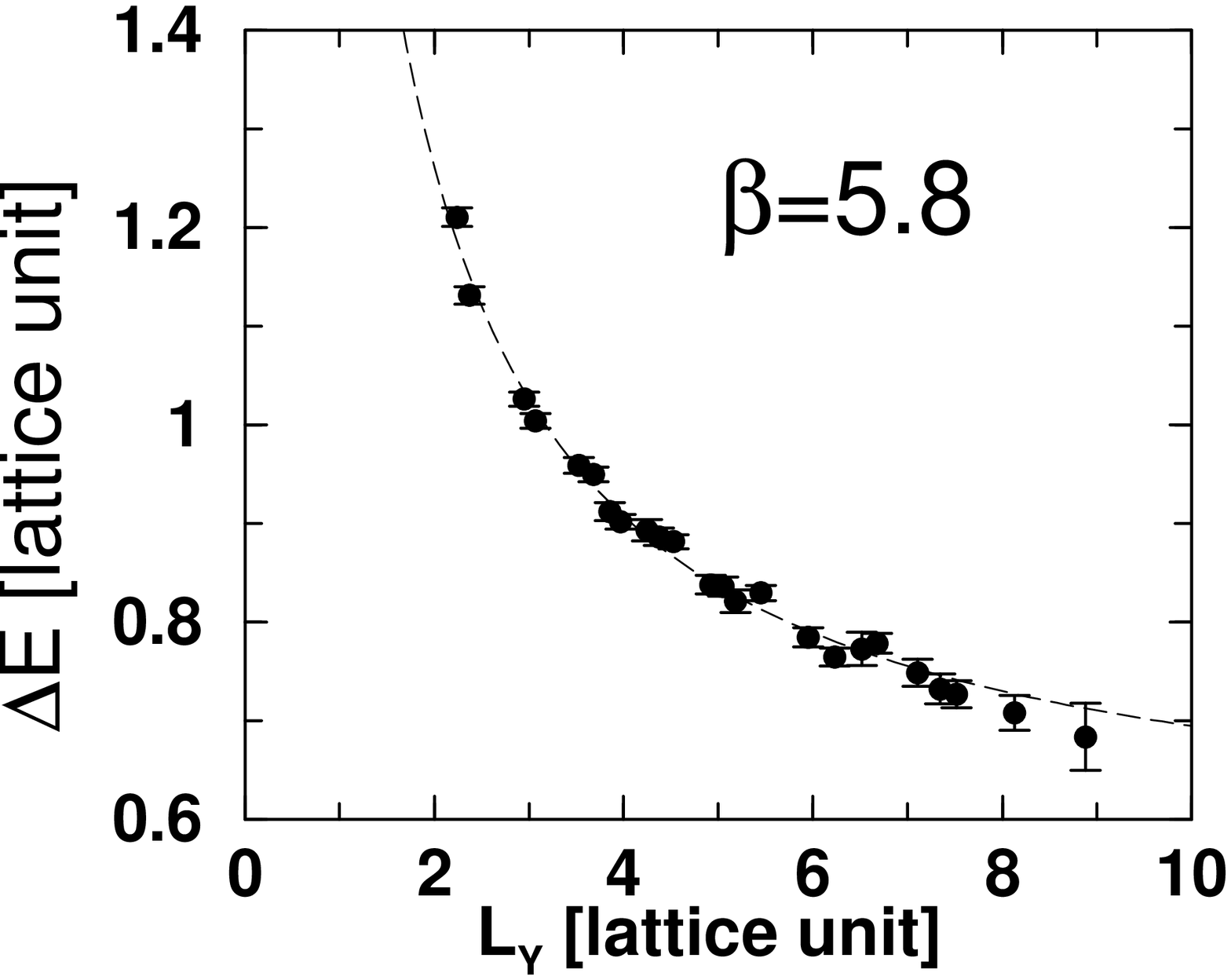}
\includegraphics[height=4.5cm]{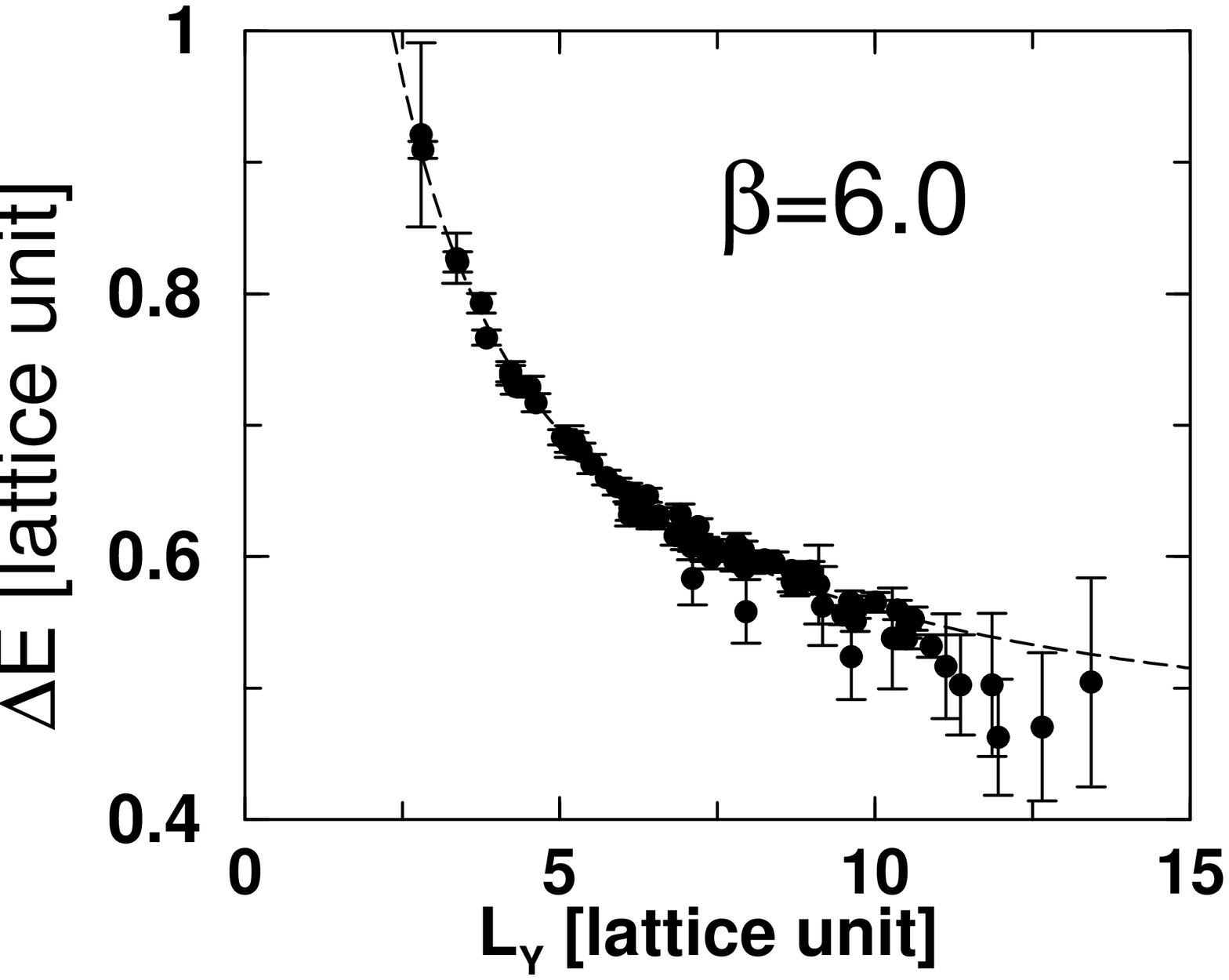}
\caption{
(a) The Mercedes form for the 3Q system. (b) \& (c) 
Lattice QCD results of the gluonic excitation energy 
$\Delta E_{\rm 3Q} \equiv V^{\rm e.s.}_{\rm 3Q}-V^{\rm g.s.}_{\rm 3Q}$
in the 3Q system plotted against the modified Y-length $L_{\overline{\rm Y}}$.
The dashed curve denotes the inverse Mercedes Ansatz. 
}
\end{center}
\end{figure}

\vspace{-0.85cm}

\subsection{Behind the Success of the Quark Model}

Here, we consider the connection between QCD and the quark model 
in terms of the gluonic excitation.\cite{TS03,TS04,STI04,STOI04} 
While QCD is described with quarks and gluons, 
the simple quark model successfully describes low-lying hadrons 
even without explicit gluonic modes.
In fact, the gluonic excitation seems invisible in low-lying hadron spectra, 
which is rather mysterious.

On this point, we find the gluonic-excitation energy to be about 1GeV or more, 
which is rather large compared with the excitation energies of the quark origin.
Therefore, the contribution of gluonic excitations 
is considered to be negligible and the dominant contribution is brought 
by quark dynamics such as the spin-orbit interaction for low-lying hadrons. 
Thus, the large gluonic-excitation energy of about 1GeV gives the physical reason for 
the invisible gluonic excitation in low-lying hadrons, 
which would play the key role for the success of the quark model 
without gluonic modes.\cite{TS03,TS04,STI04,STOI04} 

In Fig.4, 
we present a possible scenario from QCD to the massive quark model 
in terms of color confinement and dynamical chiral-symmetry breaking (DCSB).

\vspace{-0.5cm}

\begin{figure}[h]
\begin{center}
\includegraphics[height=9.5cm]{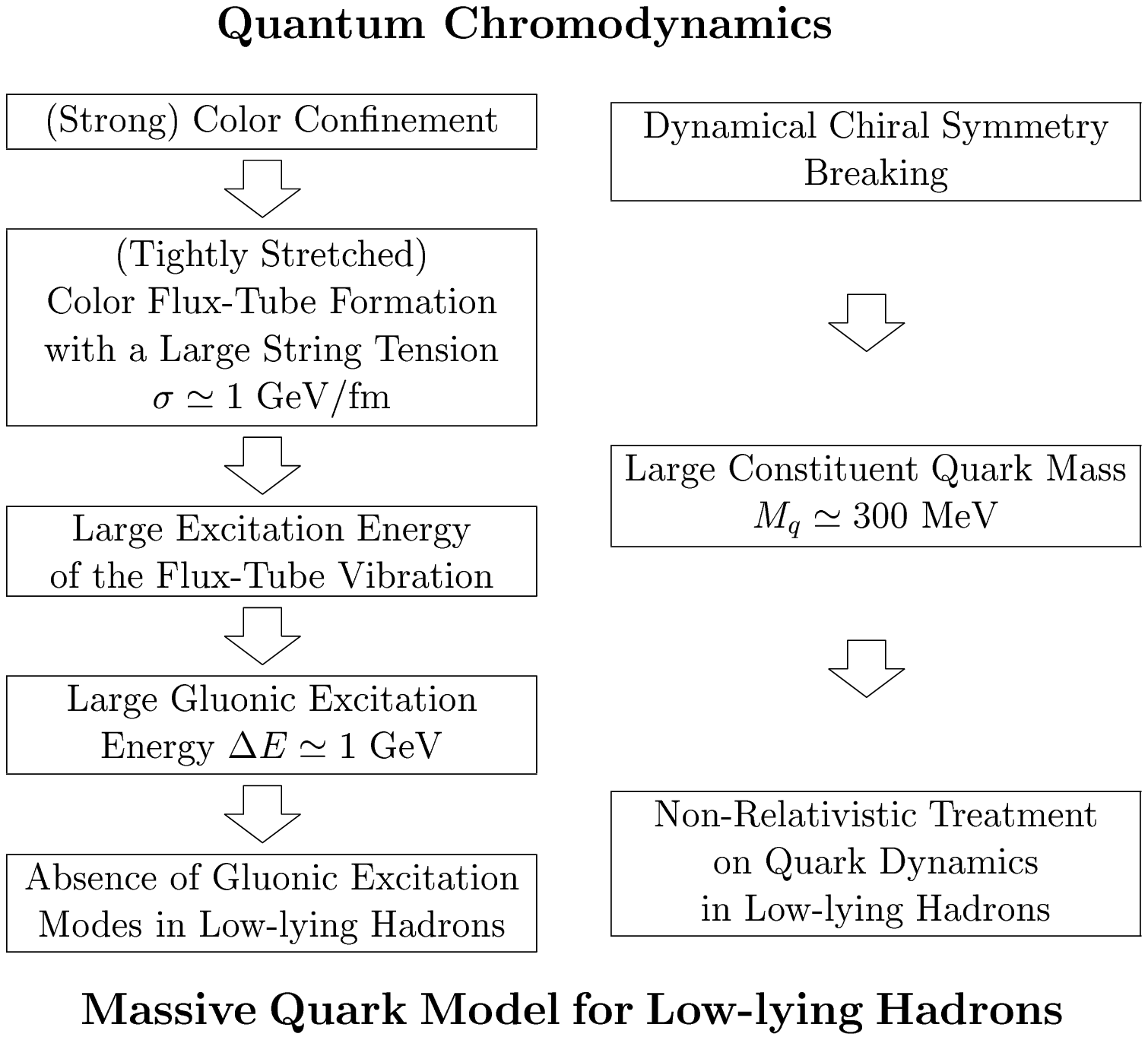}
\caption{A possible scenario from QCD to the quark model in terms of 
color confinement and DCSB.
DCSB leads to a large constituent quark mass of about 300 MeV, which enables the non-relativistic treatment 
for quark dynamics approximately. 
Color confinement results in the color flux-tube formation among quarks with a large string tension of $\sigma \simeq$ 1 GeV/fm.
In the flux-tube picture, the gluonic excitation is described as the flux-tube vibration, 
and its energy is expected to be large in the hadronic scale.
The large gluonic-excitation energy of about 1 GeV leads to 
the absence of the gluonic mode in low-lying hadrons, 
which plays the key role to the success of the quark model without gluonic excitation modes.}
\end{center}
\end{figure}

\section{Tetra-quark and Penta-quark Potentials}

In this section, we perform the first study of the multi-quark potentials in SU(3) lattice QCD, 
motivated by recent experimental discoveries of multi-quark hadrons, i.e.,  
X(3872) and $D_s(2317)$ as the candidates of tetra-quark (QQ-$\rm \bar Q \bar Q$) mesons, and 
$\Theta^+(1540)$, $\Xi^{--}(1862)$ and $\Theta_c(3099)$ as penta-quark (4Q-$\rm \bar Q$) baryons.\cite{Z04}
As the unusual features of multi-quark hadrons, their decay widths are extremely narrow, e.g., $\Gamma({\rm X}(3872)) < 2.3{\rm MeV}$ (90 \% C.L.).
For the physical understanding of multi-quark hadrons, 
theoretical analyses are necessary as well as the experimental studies. 
In particular, to clarify the inter-quark force in the multi-quark system based on QCD 
is required for the realistic model calculation of multi-quark hadrons. 

\subsection{OGE Coulomb plus Multi-Y Ansatz}

As the theoretical form of the multi-quark potential, 
we present one-gluon-exchange (OGE) Coulomb plus 
multi-Y Ansatz\cite{STOI04,OST04,OST04p} based on Eq.(1), i.e., the 
sum of OGE Coulomb potentials and the linear confinement potential 
proportional to the length $L_{\rm min}$ of the multi-Y shaped flux-tube.

On the 4Q potential $V_{\rm 4Q}$, we investigate the QQ-$\rm \bar Q\bar Q$ system where 
two quarks locate at (${\bf r}_1$, ${\bf r}_2$) and two antiquarks at (${\bf r}_3$, ${\bf r}_4$) as shown in Fig.5. 
For the connected 4Q system, the plausible form of $V_{\rm 4Q}$ is OGE plus multi-Y Ansatz,\cite{OST04p}  
\begin{equation}
V_{\rm c4Q} \equiv -A_{\rm 4Q}\{(\frac1{r_{12}}+\frac1{r_{34}})
+\frac1{2}(\frac1{r_{13}}+\frac1{r_{14}}+\frac1{r_{23}}+\frac1{r_{24}})\}
+\sigma_{\rm 4Q}L_{\rm min}+C_{\rm 4Q}, 
\end{equation}
while $V_{\rm 4Q}$ for the disconnected 4Q system would be approximated by 
the ``two-meson" Ansatz as $V_{\rm 2Q\bar Q} \equiv V_{\rm Q\bar Q}(r_{13})+V_{\rm Q\bar Q}(r_{24})$.

On the 5Q potential $V_{\rm 5Q}$, we investigate the QQ-$\rm \bar Q$-QQ system  
where the two quarks at (${\bf r}_1$, ${\bf r}_2$) and those 
at (${\bf r}_3$, ${\bf r}_4$) form $\bar {\bf 3}$ representation of SU(3) color, respectively, 
and the antiquark locates at ${\bf r}_5$, as shown in Fig.5. 
For the 5Q system, OGE Coulomb plus multi-Y Ansatz is expressed as $V_{\rm 5Q}=V_{\rm 5Q}^{\rm Coul}
+\sigma_{\rm 5Q} L_{\rm min}+C_{\rm 5Q}$ with the Coulomb part as 
\begin{eqnarray}
V_{\rm 5Q}^{\rm Coul}=&-&A_{\rm 5Q}\{ ( \frac1{r_{12}}  + \frac1{r_{34}}) 
+\frac12(\frac1{r_{15}} +\frac1{r_{25}} +\frac1{r_{35}} +\frac1{r_{45}}) \nonumber \\
&+&\frac14(\frac1{r_{13}} +\frac1{r_{14}} +\frac1{r_{23}} +\frac1{r_{24}}) \}.
\end{eqnarray}

We theoretically set $(A_{\rm 4Q},\sigma_{\rm 4Q})$ and $(A_{\rm 5Q},\sigma_{\rm 5Q})$ 
to be $(A_{\rm 3Q},\sigma_{\rm 3Q}) \simeq (0.1366, 0.046a^{-2})$ 
in the 3Q potential.\cite{TSNM02}
Note that there is no adjustable parameter in the theoretical Ans\"atze 
except for an irrelevant constant.

\begin{figure}[h]
\begin{center}
\includegraphics[height=2.3cm]{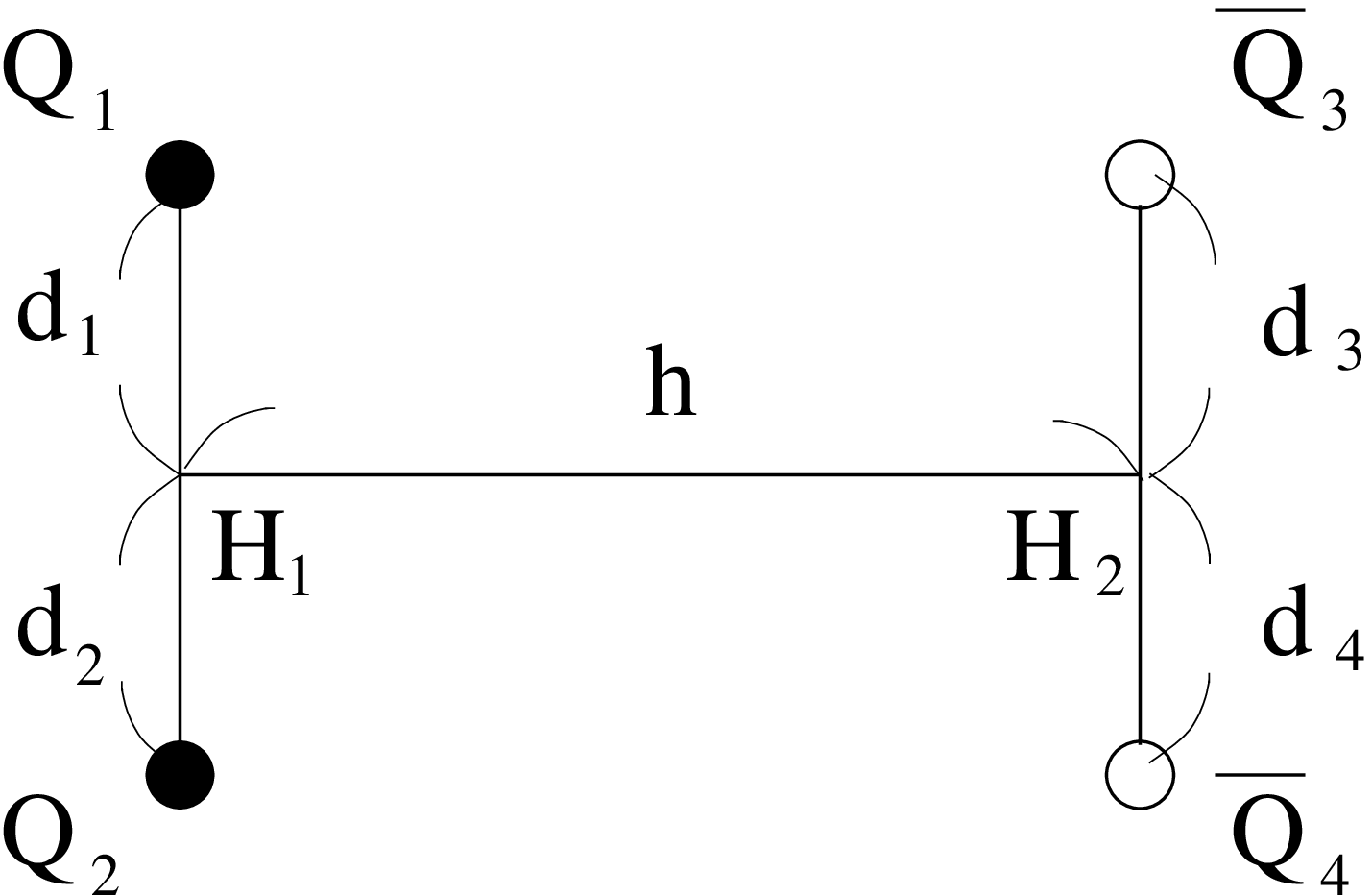} 
\includegraphics[height=2.3cm]{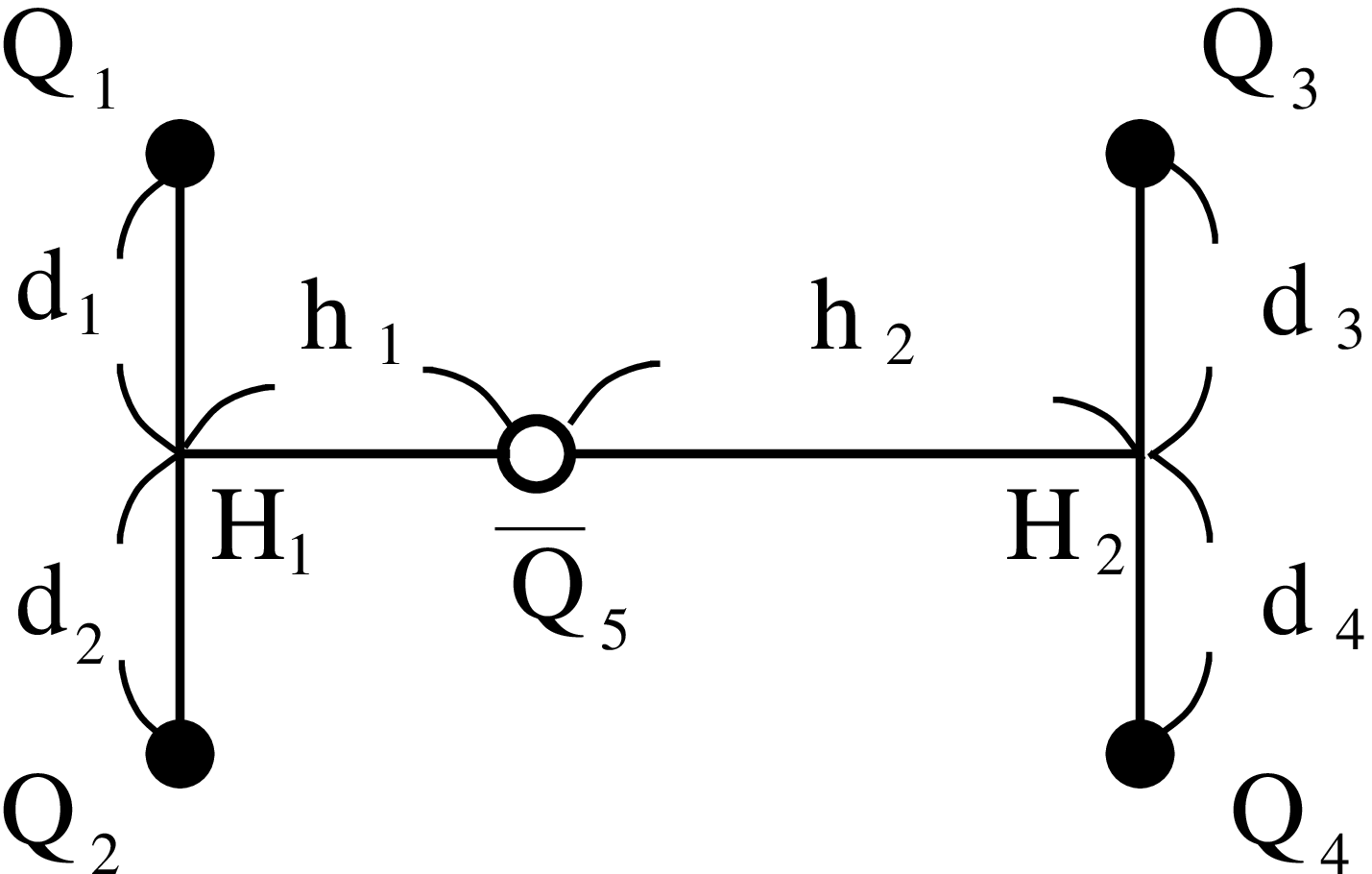} 
\includegraphics[height=2.3cm]{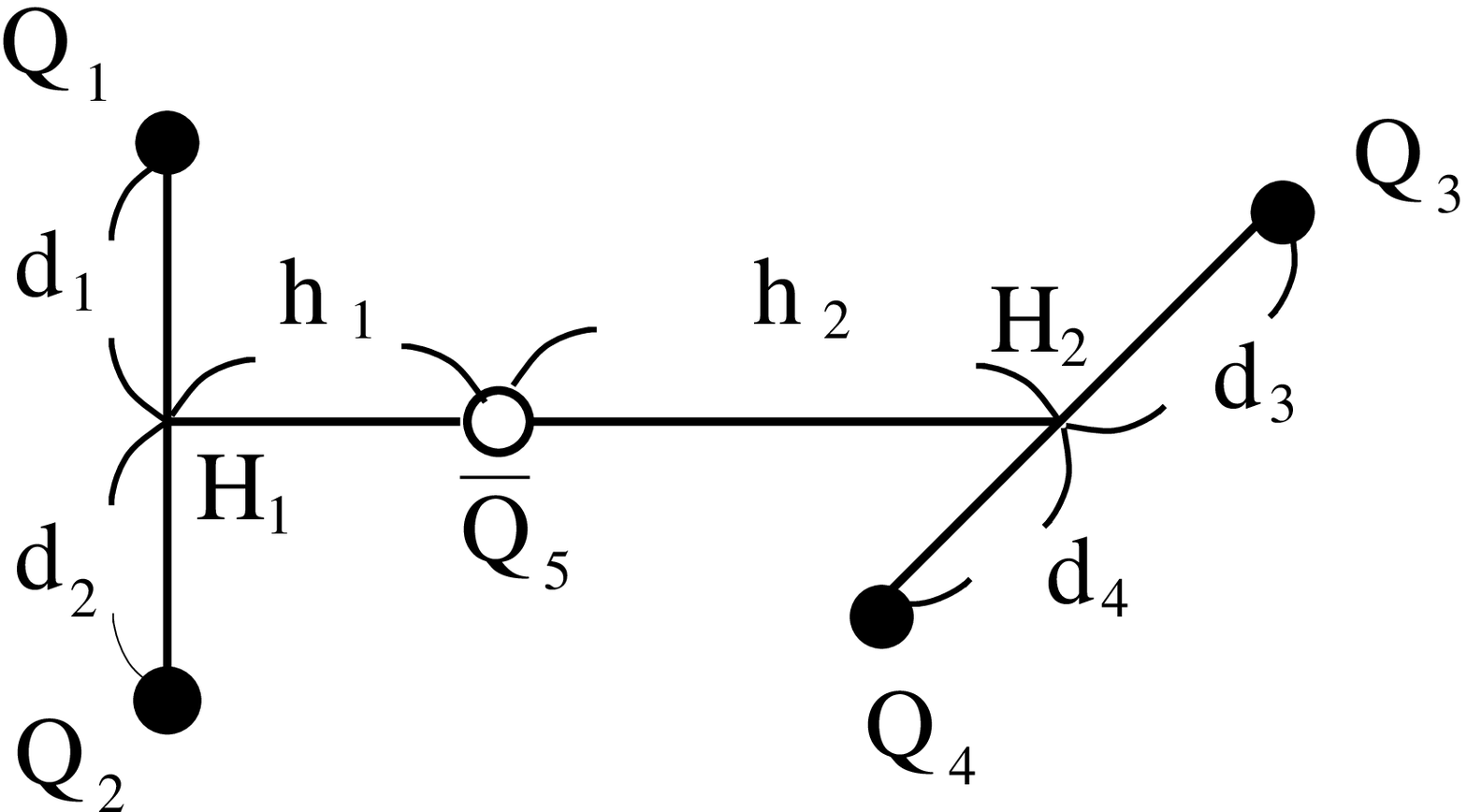} 
\caption{
(a) A planar tetra-quark (QQ-$\rm \bar Q\bar Q$) configuration.
(b) A planar penta-quark (4Q-$\rm \bar Q$) configuration.
(c) A twisted penta-quark configuration with  
${\rm Q}_1{\rm Q}_2 \perp {\rm Q}_3{\rm Q}_4$.
We here take $d_1=d_2=d_3=d_4\equiv d$.}
\end{center}
\end{figure}

\subsection{Multi-quark Wilson loops and multi-quark potentials}

In QCD, the static multi-quark potentials can be obtained from the corresponding multi-quark Wilson loops.  
As shown in Fig.6, we define the 4Q Wilson loop $W_{\rm 4Q}$ and 
the 5Q Wilson loop $W_{\rm 5Q}$\cite{STOI04,OST04,OST04p} by  
\begin{eqnarray}
W_{\rm 4Q} &\equiv& \frac1{3} {\rm tr}
(\tilde M_1\tilde L_{12}\tilde M_2\tilde R_{12}), \nonumber \\
W_{\rm 5Q} &\equiv& \frac1{3!} \epsilon^{abc} \epsilon^{a'b'c'}
\tilde M^{aa'}
(\tilde L_3\tilde L_{12}\tilde L_4)^{bb'}
(\tilde R_3\tilde R_{12}\tilde R_4)^{cc'}, 
\end{eqnarray}
where $\tilde L_i, \tilde R_i, \tilde M, \tilde M_j \; (i=1,2,3,4, j=1,2)$ are given by 
\begin{equation}
\tilde L_i, \tilde R_i, \tilde M, \tilde M_j 
\equiv P\exp\{ig \int_{L_i, R_i, M, M_j} dx^{\mu}A_{\mu}(x)\} \in {\rm SU(3)}_c,
\end{equation}
i.e.,  
$\tilde L_i, \tilde R_i, \tilde M, \tilde M_j\; (i=3,4, j=1,2)$ 
are line-like variables and 
$\tilde L_i, \tilde R_i\; (i=1,2)$ are staple-like variables, 
and $\tilde L_{12}, \tilde R_{12}$ are defined by  
\begin{equation}
\tilde L_{12}^{a'a}\equiv \frac12 \epsilon^{abc} \epsilon^{a'b'c'}
\tilde L_1^{bb'} \tilde L_2^{cc'},
\hspace{0.2cm}
\tilde R_{12}^{a'a}\equiv \frac12 \epsilon^{abc} \epsilon^{a'b'c'}
\tilde R_1^{bb'} \tilde R_2^{cc'}. 
\end{equation}
Note that both the 4Q Wilson loop $W_{\rm 4Q}$ and the 5Q Wilson loop $W_{\rm 5Q}$ are gauge invariant.

\begin{figure}[h]
\begin{center}
\includegraphics[height=2.8cm]{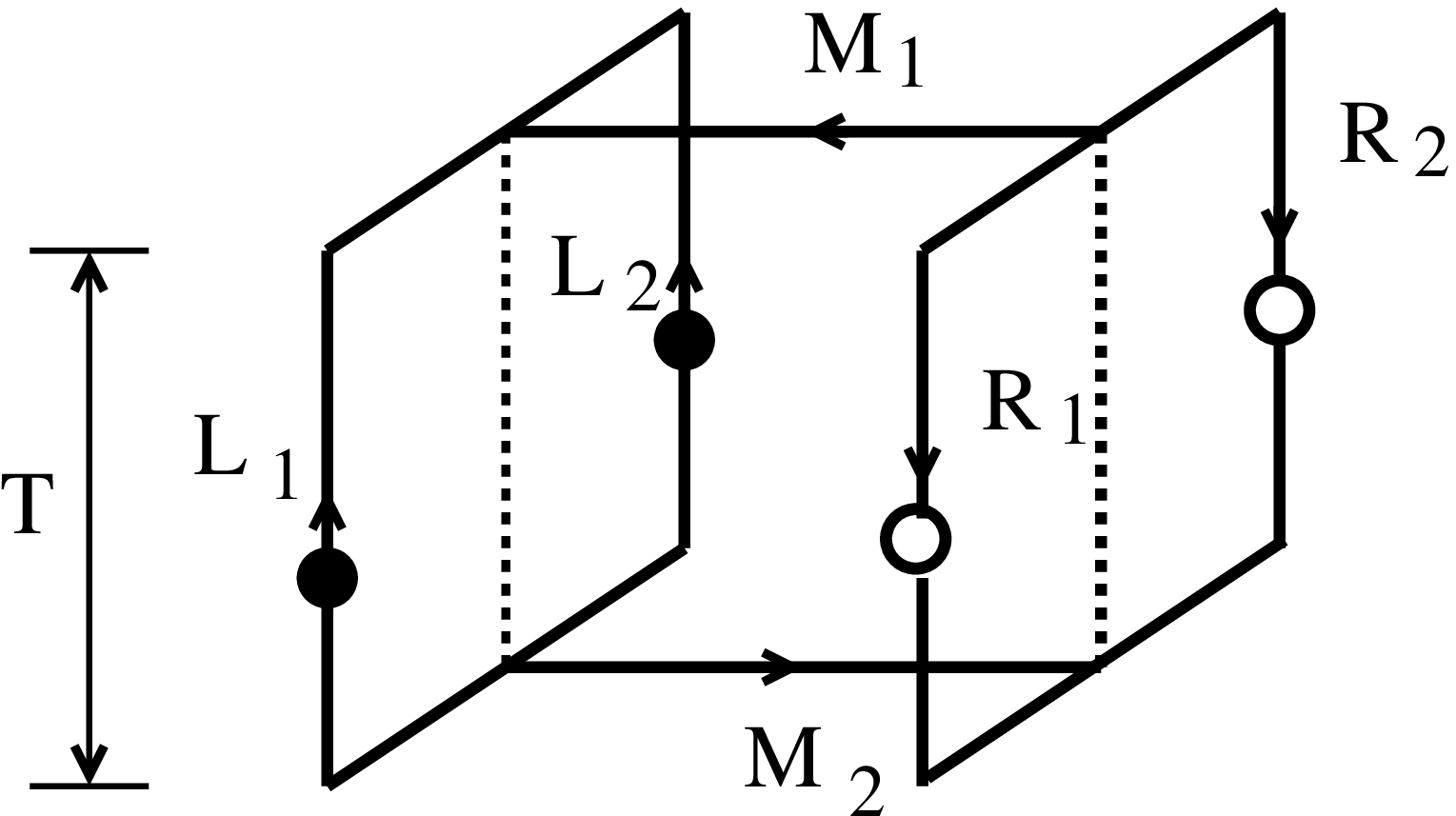} 
\includegraphics[height=2.8cm]{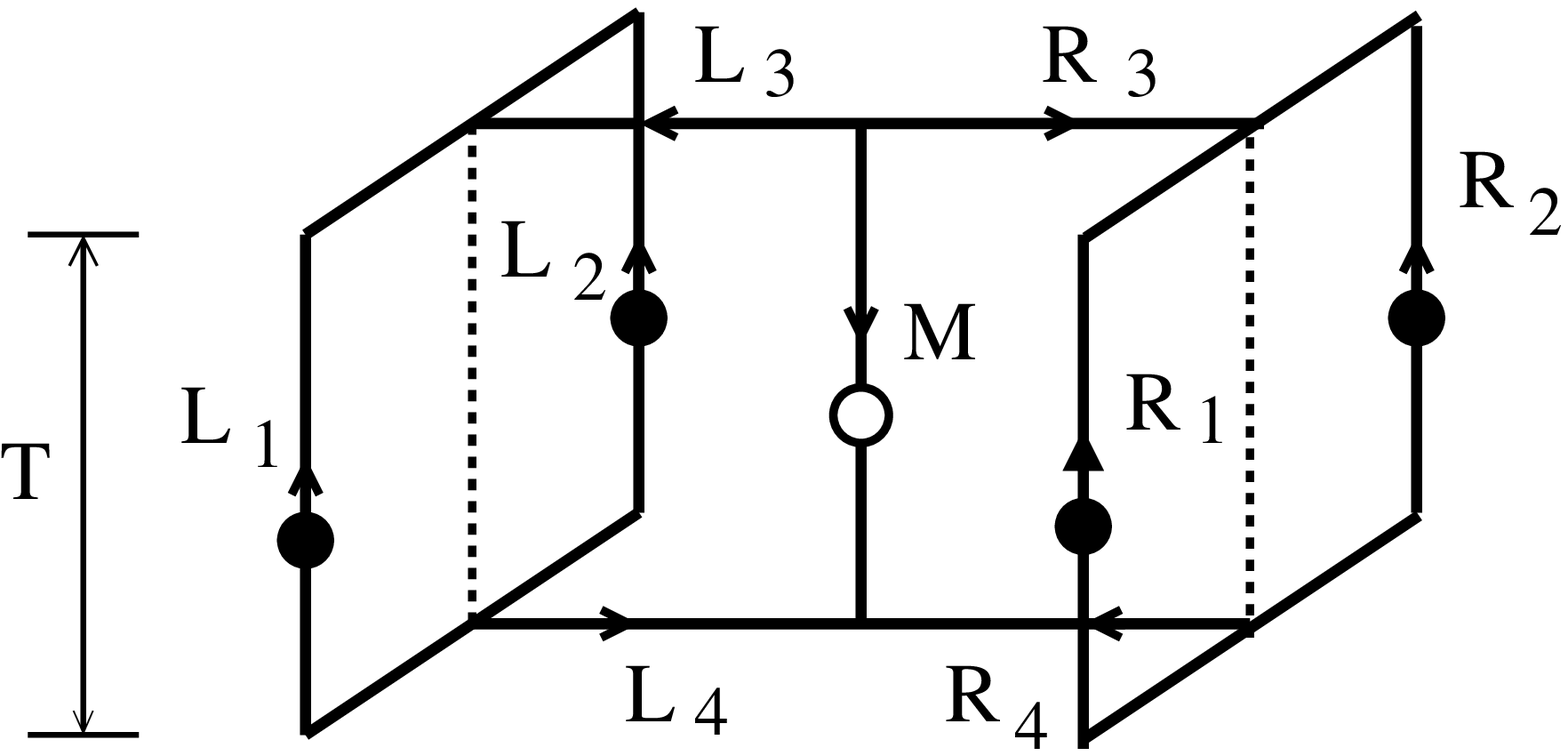} 
\caption{
(a) The tetra-quark (4Q) Wilson loop $W_{\rm 4Q}$. 
(b) The penta-quark (5Q) Wilson loop $W_{\rm 5Q}$. 
}
\end{center}
\end{figure}

We calculate the multi-quark potentials ($V_{\rm 4Q}$, $V_{\rm 5Q}$)
from the multi-quark Wilson loops ($W_{\rm 4Q}$, $W_{\rm 5Q}$) in SU(3) lattice QCD 
with $\beta=6.0$ (i.e., $a \simeq 0.1{\rm fm}$) and $16^3 \times 32$ at the quenched level,\cite{STOI04,OST04,OST04p}
using the smearing method to reduce the excited-state components.
In this paper, we investigate the planar and twisted configurations 
for the multi-quark system as shown in Fig.5, 
and show the results for $d_1=d_2=d_3=d_4 \equiv d$ and $h_1=h_2 \equiv h/2$.

Figure 7 shows the 4Q potential $V_{\rm 4Q}$.\cite{OST04p}
For large $h$, $V_{\rm 4Q}$ coincides with the energy $V_{\rm c4Q}(d,h)$ of the connected 4Q system. 
For small $h$, $V_{\rm 4Q}$ coincides with the energy $V_{\rm 2Q\bar Q}=2V_{\rm Q\bar Q}(h)$ 
of the ``two-meson" system composed of two flux-tubes.
Thus, we get the relation of $V_{\rm 4Q} = {\rm min} \{V_{\rm c4Q}(d,h), 2V_{\rm Q\bar Q}(h)\}$, 
and find the ``flip-flop" between the connected 4Q system and the ``two-meson" system around 
the level-crossing point where these two systems are degenerate as $V_{\rm c4Q}(d,h)=2V_{\rm Q\bar Q}(h)$. 

\begin{figure}
\begin{center}
\includegraphics[height=5.5cm,angle=270]{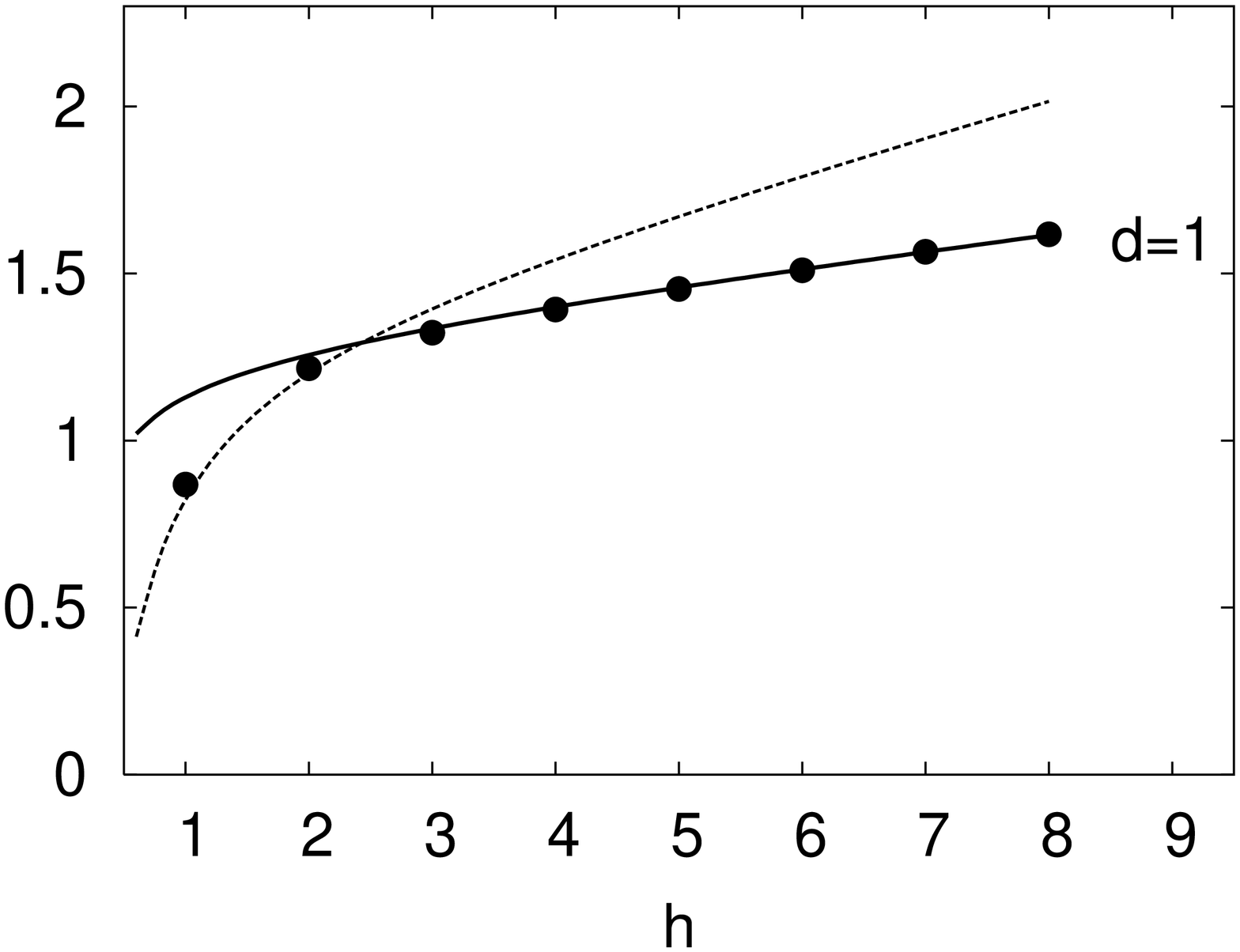}
\includegraphics[height=5.5cm,angle=270]{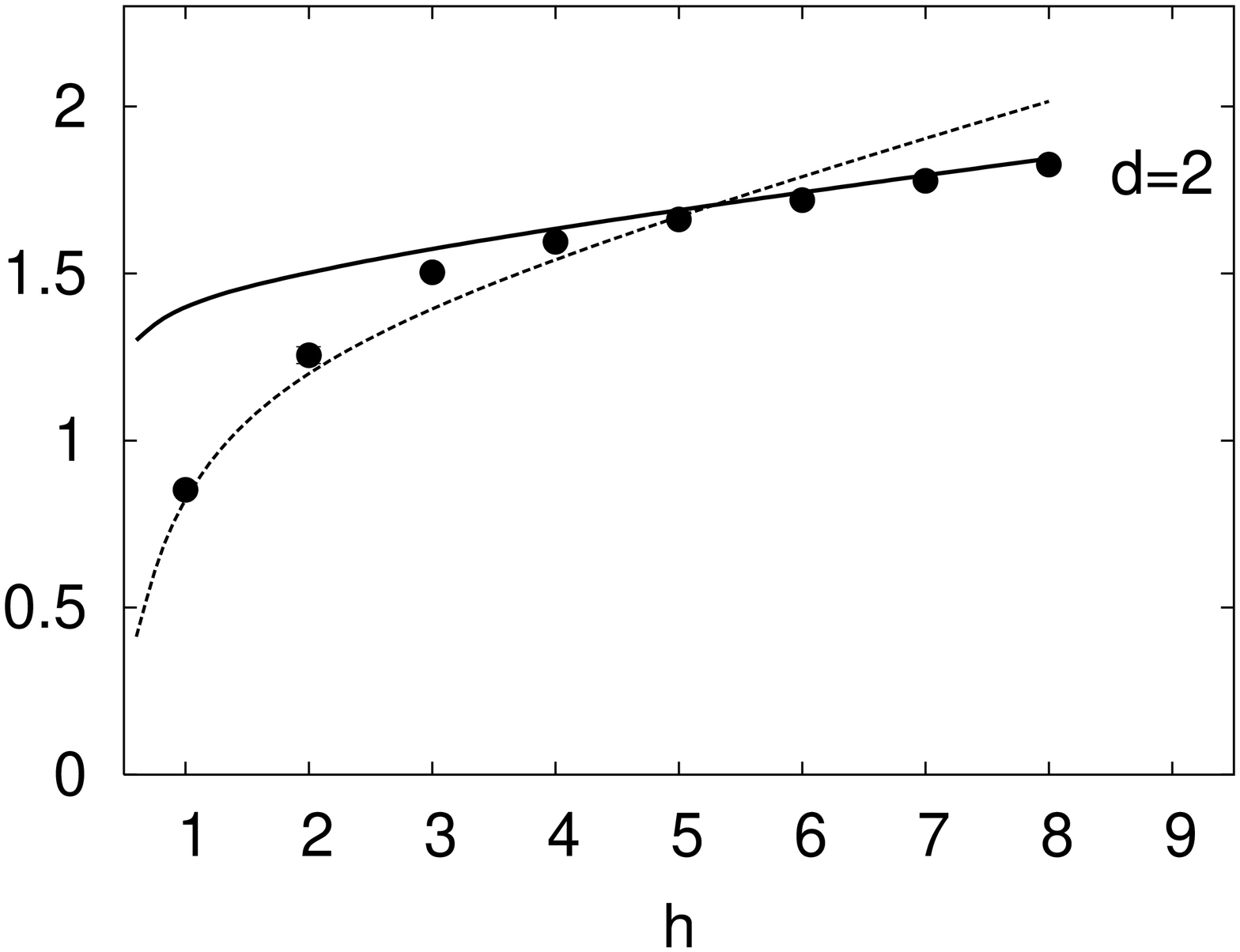}
\caption{The 4Q potential $V_{\rm 4Q}$ in the lattice unit 
for the planar 4Q configuration 
with $d=1$ (left) and $d=2$ (right) as shown in Fig.5(a). 
The symbols denote the lattice QCD results.
The theoretical curves are added for the connected 4Q system (the solid curve) 
and for the ``two-meson" system (the dashed curve).} 
\end{center}
\end{figure}

In Fig.8, we show the 5Q potential $V_{\rm 5Q}$. 
The lattice data denoted by the symbols are found to be well reproduced by 
the theoretical curve of the OGE plus multi-Y Ansatz\cite{STOI04,OST04,OST04p} 
with $(A_{\rm 5Q},\sigma_{\rm 5Q})$ fixed to be $(A_{\rm 3Q},\sigma_{\rm 3Q})$ in the 3Q potential.\cite{TSNM02}

\begin{figure}[h]
\begin{center}
\includegraphics[height=3.9cm]{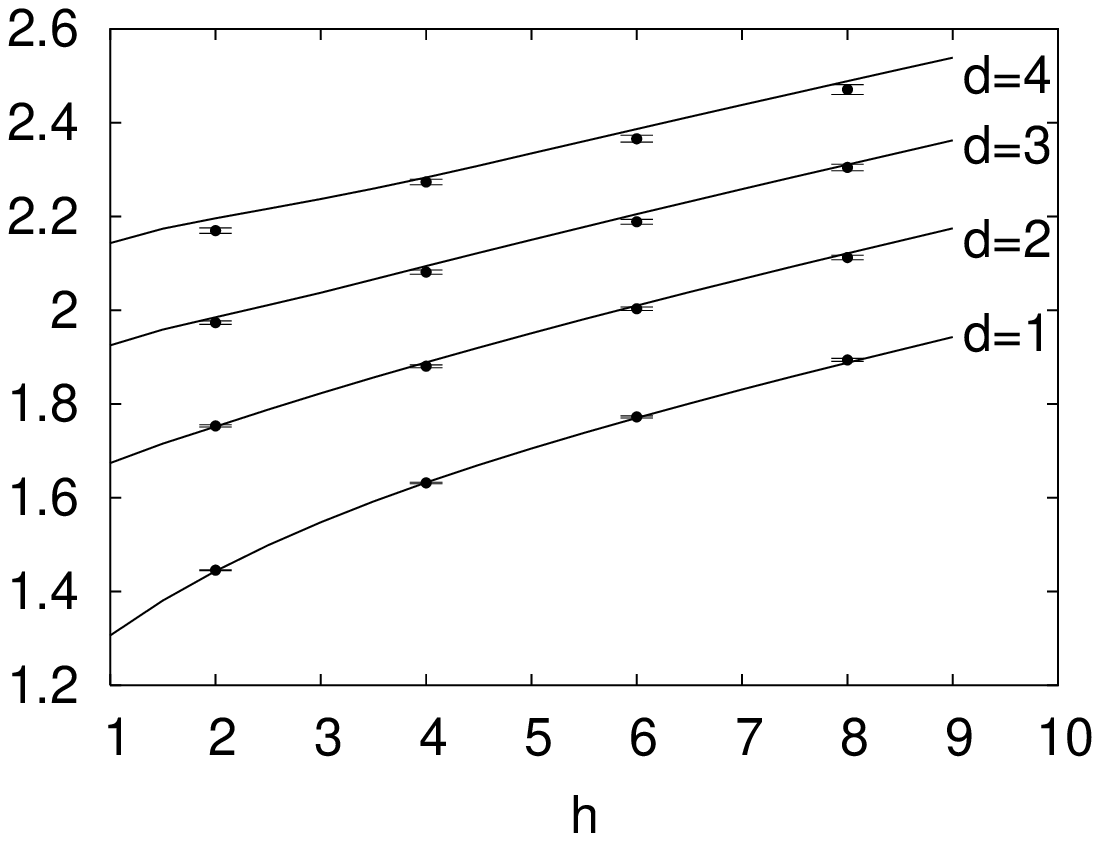}
\includegraphics[height=3.9cm]{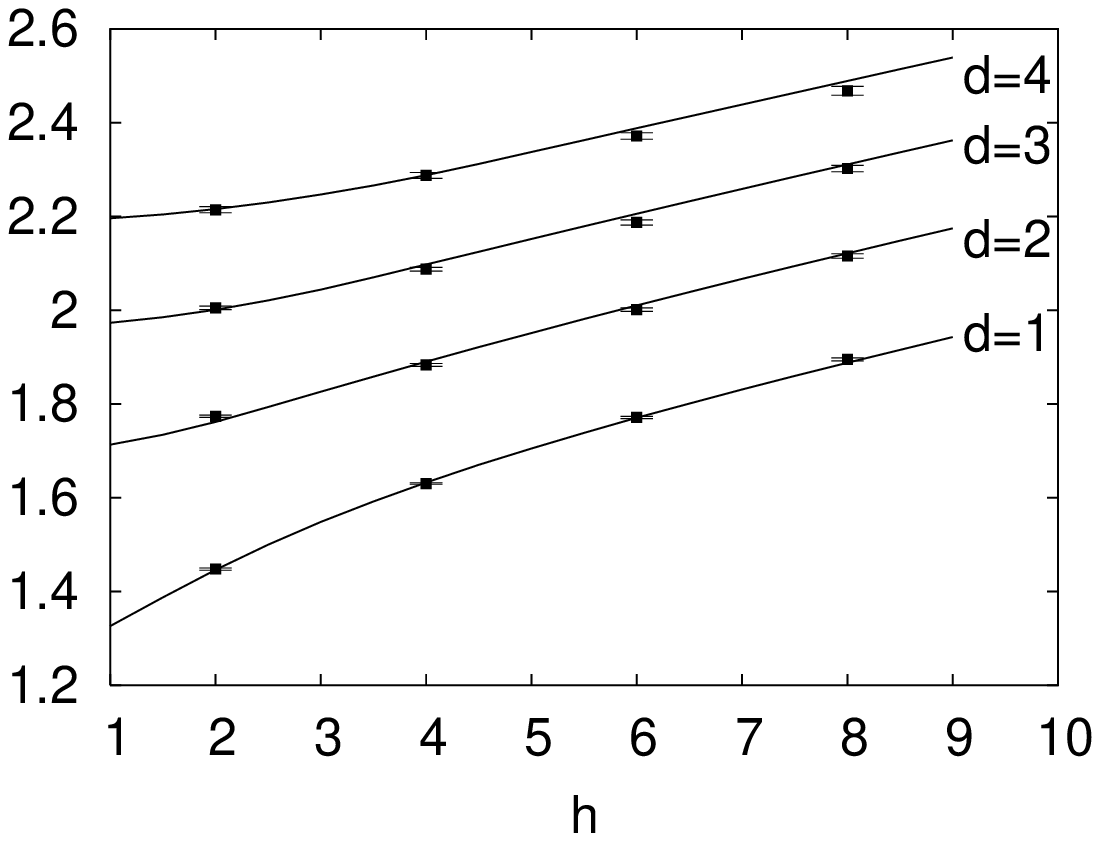} 
\caption{The 5Q potential $V_{\rm 5Q}$ in the lattice unit 
for the planar configurations (left) and the twisted configurations (right) as shown in Figs.5(b) and (c). 
The symbols denote the lattice QCD results. We add the theoretical curve 
of the OGE plus multi-Y Ansatz with $(A_{\rm 5Q},\sigma_{\rm 5Q})$ fixed to be $(A_{\rm 3Q},\sigma_{\rm 3Q})$.}
\end{center}
\end{figure}

As a remarkable fact, we find the universality of the string tension 
and the OGE result among Q$\bar {\rm Q}$, 3Q, 4Q and 5Q systems 
as\cite{TMNS01,TSNM02,TS03,TS04,STI04,STOI04,OST04,OST04p}
\begin{eqnarray}
\sigma_{\rm Q\bar{\rm Q}}\simeq \sigma_{\rm 3Q} \simeq \sigma_{\rm 4Q} 
\simeq \sigma_{\rm 5Q}, \quad
\frac12A_{\rm Q\bar{\rm Q}}\simeq A_{\rm 3Q} \simeq A_{\rm 4Q} \simeq A_{\rm 5Q}.
\end{eqnarray} 
This result supports the flux-tube picture 
on the confinement mechanism even for the multi-quark system.\cite{STOI04,OST04,OST04p}

\section{QCD String Theory for the Penta-Quark Decay}

Our lattice QCD studies on the various inter-quark potentials 
indicate the flux-tube picture for hadrons, 
which is idealized as the QCD string model.
In this section, we consider penta-quark dynamics, 
especially for its extremely narrow width, in terms of the QCD string theory.

The ordinary string theory mainly describes open and closed strings corresponding to mesons and glueballs, 
and has only two types of the reaction process as shown in Fig.9:
\begin{enumerate}
\item[1.] The string breaking (or fusion) process.
\item[2.] The string recombination process.
\end{enumerate}

\vspace{-0.5cm}

\begin{figure}[h]
\begin{center}
\includegraphics[height=4.5cm]{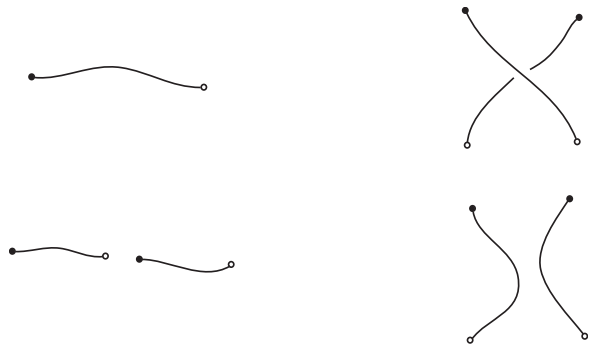}
\caption{The reaction process in the ordinary string theory:  
the string breaking (or fusion) process (left) and 
the string recombination process (right).}
\end{center}
\end{figure}

On the other hand, the QCD string theory describes also 
baryons and antibaryons as the Y-shaped flux-tube, which  is 
different from the ordinary string theory.
Note that the appearance of the Y-type junction is peculiar to the QCD string theory.
Accordingly, the QCD string theory includes the third reaction process as shown in Fig.10:
\begin{enumerate}
\item[3.] The junction (J) and anti-junction ($\rm \bar J$) par creation (or annihilation) process.
\end{enumerate}
\begin{figure}[h]
\begin{center}
\includegraphics[width=10cm]{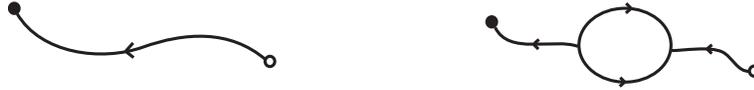}
\caption{
The junction (J) and anti-junction ($\rm \bar J$) par creation (or annihilation) process peculiar to the QCD string theory.}
\end{center}
\end{figure}
Through this J-$\bar {\rm J}$ pair creation process, 
the baryon and anti-baryon pair creation can be described.

As a remarkable fact in the QCD string theory, 
the decay process (or the creation process) of penta-quark baryons inevitably 
accompanies the J-$\bar {\rm J}$ creation\cite{BKST04} 
as shown in Fig.11.
Here, the intermediate state is considered as a gluonic-excited state, since it clearly corresponds to 
a non-quark-origin excitation.

Our lattice QCD study indicates that 
such a gluonic-excited state is a highly-excited state with the excitation energy above 1GeV.
Then, in the QCD string theory, 
the decay process of the penta-quark baryon near the threshold 
can be regarded as a quantum tunneling, 
and therefore the penta-quark decay is expected to be strongly suppressed.
This leads to a very small decay width of penta-quark baryons.

\begin{figure}[h]
\begin{center}
\includegraphics[width=11.5cm]{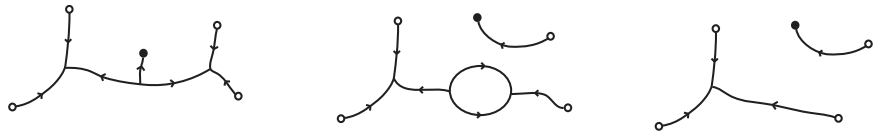}
\caption
{A decay process of the penta-quark baryon in the QCD string theory.
The penta-quark decay process inevitably accompanies the J-$\bar {\rm J}$ creation, which is a kind of the gluonic excitation.
}
\end{center}
\end{figure}

Now, we try to estimate the decay width of penta-quark baryons in the QCD string theory.
In the quantum tunneling as shown in Fig.11, 
the barrier height corresponds to the gluonic excitation energy $\Delta E$ of the intermediate state, 
and can be estimated as $\Delta E \simeq$ 1GeV. 
The time scale $T$ for the tunneling precess is expected to be the hadronic scale 
as $T =0.5 \sim 1{\rm fm}$, since $T$ cannot be smaller than the spatial size of the reaction area   
due to the causality.
Then, the suppression factor for the penta-quark decay can be roughly estimated as 
\begin{eqnarray}
|\exp(-\Delta E T)|^2 \simeq |e^{-1{\rm GeV} \times (0.5 \sim 1){\rm fm}}|^2 \simeq 
10^{-2}\sim 10^{-4}.
\end{eqnarray}
Note that this suppression factor $|\exp(-\Delta E T)|^2$
appears in the decay process of the penta-quark baryons 
for both positive-parity and negative-parity states. 

For the decay of $\Theta^+(1540)$ into N and K, the Q-value $Q$ is  
$Q= M(\Theta^+)-M({\rm N})-M({\rm K}) 
\simeq (1540-940-500) {\rm MeV} \simeq 100 {\rm MeV}$.
In ordinary sense, the decay width is expected to be controlled by  
$\Gamma_{\rm hadron} \simeq Q \simeq 100{\rm MeV}$.
Considering the extra suppression factor of $|\exp(-\Delta E T)|^2$, 
we get a rough order estimate for the decay width of $\Theta^+(1540)$ as 
$\Gamma[\Theta^+(1540)] \simeq \Gamma_{\rm hadron} \times |\exp(-\Delta E T)|^2 \simeq 1 \sim 10^{-2}$MeV.

\vspace{0.5cm}

\noindent
{\bf Acknowledgements}

\vspace{0.3cm}

\noindent
H.S. would like to thank Profs. G.M.~Prosperi and N.~Brambilla for their kind hospitality at Confinement VI.
H.S. is also grateful to Profs. T.~Kugo and A.~Sugamoto for useful discussions on the QCD string theory. 
The lattice QCD Monte Carlo simulations have been performed 
on NEC-SX5 at Osaka University and on HITACHI-SR8000 at KEK.

\end{document}